\newcommand{\Cint}[1]{\oint\frac{d#1}{2 \pi i}\;}
\newcommand{\Lop}{{\cal L}}

\newcommand{\evOper}{transfer operator}
\newcommand{\EvOper}{Transfer operator}
\newcommand{\dzeta}{dynamical zeta function}

\newcommand{\Fd}{Fredholm determinant}

\newcommand{\inFix}[1]{{\in \mbox{\footnotesize Fix}(f^{#1})}}

\newcommand{\rf}[1]{~\cite{#1}}
\newcommand{\reffig} [1] {figure~\ref{#1}}

\newcommand{\refref} [1] {ref.~\cite{#1}}

\newcommand{\refrefs}[1] {refs.~\cite{#1}}
\newcommand{\reftab} [1] {table~\ref{#1}}
\newcommand{\refsect}[1] {section~\ref{#1}}

\newcommand{\barr}{\begin{array}}
\newcommand{\ear}{\end{array}}
\newcommand{\bea}{\begin{eqnarray}}
\newcommand{\nnu}{\nonumber}
\newcommand{\eea}{\end{eqnarray}}
\newcommand{\beq}{\begin{equation}}
\newcommand{\continue}{\nonumber \\ }

\newcommand{\eeq}{\end{equation}}
\newcommand{\ee}[1] {\label{#1} \end{equation}}
\newcommand{\refeq}[1]{(\ref{#1})}

\newcommand{\MatrixII}[4]{
   \left[
   \begin{array}{cc}
      {#1}  &  {#2}  \\ [1ex]
      {#3}  &  {#4}
   \end{array}
   \right] }

\newcommand{\MatrixIII}[9]{
   \left[
   \begin{array}{ccc}
      {#1}  &  {#2} &  {#3}  \\ [1ex]
      {#4}  &  {#5} &  {#6}  \\ [1ex]
      {#7}  &  {#8} &  {#9}
   \end{array}
   \right] }

\newcommand{\combinatorial}[2]{
   \left( #1
        \atop
          #2 \right) }

\newtheorem{rmark}{{\small\bf\sf Remark}} 

\documentstyle[epsfig]{ioplppt}  
\begin{document}
\renewcommand{\baselinestretch}{1.2}
\jl{8}  
\title{Beyond the periodic orbit theory}[Beyond periodic orbits]
\author{Predrag Cvitanovi\'c, 
        Kim Hansen, Juri Rolf  and G\'abor Vattay\ftnote{2}{
Permanent address:
Dept. Solid State Physics,
E\"otv\"os University,
Muzeum krt. 6-8,
H-1088 Budapest                                         }
        }
\address{Center for Chaos and Turbulence Studies\\
Niels Bohr Institute\\
Blegdamsvej 17, DK-2100  Copenhagen \O, Denmark}
\date{\today}

\begin{abstract}
The global constraints on chaotic dynamics induced by the
analyticity of smooth flows are used to dispense with
individual periodic orbits and derive infinite families of
exact sum rules for several simple dynamical systems.
The associated \Fd s are of particularly simple polynomial form. 
The theory developed 
suggests an alternative to the conventional periodic orbit theory approach
to determining eigenspectra of \evOper s.
\end{abstract}
\pacs{0320, 0365, 0545}


\vskip 10mm 
\hfill {\em Dedicated to Adrian Douady for his 60th birthday}\\

\section{Introduction}

Low dimensional chaotic classical and quantum dynamical
systems\rf{ruelle,gutbook} 
can be analyzed in terms of unstable periodic orbits. The periodic
orbit theory of such systems
has been successfully applied to a wide range of physical
problems\rf{CHAOS92,CHAOS93,CC94}. However, since the phase space is
tessellated into linearized neighbourhoods of periodic points,
analyticity properties of smooth flows are not fully utilized
in the conventional periodic orbit theory.

In this paper we explore global constraints on chaotic dynamics
induced by
the analyticity of the flow. We propose to dispense
with the periodic {\em orbits} altogether,
and extract spectra of \evOper s from dynamics-induced relations
among periodic orbit {\em sums}.

For  example, evaluating trace formulas for 2-dimensional Hamiltonian 
flows\rf{dynamo} we have found exact  periodic orbit sum rules
\bea
 \sum_{i\inFix{n}} 
   \frac{1}{\Lambda_i(1-1/\Lambda_i)^2} 
        &=& 0 
\continue
 \sum_{i\inFix{n}}
   \frac{\Lambda_i+1/\Lambda_i}{\Lambda_i(1-1/\Lambda_i)^2}
        &=& 2^n \,,
\nnu
\eea
where $f$ is a volume preserving H\'enon map and
$\Lambda_i$ is the expanding eigenvalue of its $i$th 
periodic point, real or complex.
Such relations are remarkable insofar they
require a high degree of correlation  between the stabilities 
of exponentially large numbers of 
periodic orbits distributed over the entire phase space.  
As a matter of fact, for at least two of the dynamical systems studied here,
the quadratic polynomial map and the Farey map, the information
carried by the periodic points
is so redundant, that the periodic points must satisfy {\em infinitely many}
independent sum rules.

The key idea behind the method by which the sum rules are
derived is to replace a weighted sum over periodic orbits 
by an integral representation; the dynamics then induces recursion relations
between these for cycles of different topological length. 
In posing the problem this way we
draw inspiration from the work of
Eremenko, Levin,  Sodin, and  Yuditskii\rf{Levin89,Levin92,Levin93} 
and Hatjispiros and Vivaldi\rf{HV93} on the polynomial mappings, and 
Contucci and Knauf\rf{CK95} for the Farey map.
For these models the recursions relate finite numbers of terms,
and yield expressions \refeq{Fred_B}, \refeq{Farey62a}
for the corresponding \Fd s 
in terms of {\em finite} polynomials of form
\[
 \det(1-z\Lop) = \det\left(1-z {\bf L}\right)
\]
where $\Lop$ is a weighted \evOper\, and ${\bf L}$ is the
corresponding finite [$\ell$$\times$$\ell$] dimensional matrix that 
relates $\ell$ consecutive sums.

While this finiteness is a very special property of the particularly
simple models considered and cannot hold for a generic dynamical flow,
the method nevertheless suggests that one might be able to
dispense with the periodic orbit theory altogether, and
manipulate instead the traces of dynamical \evOper s by 
means of such integral representations. This tempting
possibility motivates us to explore the applicability of the method
in some depth, the main results being the sum rules and
closed form expressions for \Fd s of weighted \evOper s.

We start with polynomial mappings in \refsect{s_Contour_int_sums},
and then specialize to quadratic mappings (\refsect{s_tr_3}) for
which an infinity of sum rules is easily derived. Special examples
are worked out in the appendices.
In \refsect{s_Julia_esc} we generalize a result of \refref{Levin92}
about the spectrum of a non-polynomial \evOper for quadratic mappings.
In \refsect{s_Mult-dim} we show that our method can be extended to
higher dimensions and apply it to the H\'enon map
and the kinematic dynamo model.
Finally, in order to show that the contour
integrals are not the essence of the method, 
in \refsect{s_CIRC} we derive the same kind of explicit polynomial \Fd s for
the circle-map (or spin-chain) thermodynamics.
We discuss the applicability of the method to continuous flows
in \refsect{s_Norm_forms}: 
the sum rules offer a new invariant characterization of 
the errors due to approximating a flow by a polynomial Poincar\'e
section return mapping.
The sum rules of \refsect{s_tr_3}
use a ``signed'' rather than the ``natural'' measure;
in \refsect{s_period-doub} we outline how they might be applicable to 
the period-doubling presentation function, for which the
``signed'' measure yields a formula for the Feigenbaum $\delta$.
We finish with a critical summary and relegate
some details to appendices.

\section{\EvOper s and \Fd s}

We start with a brief summary of the dynamical systems concepts 
needed in what follows. For motivation and background we refer the reader to 
literature, for example \refrefs{CRR93,cycl_book}.

A dynamical flow is usually investigated  via  its Poincar\'e
return map $f$, with the 
evolution of a density function $\phi(x)$ given by 
\begin{equation}
  \label{top}
  (\Lop \phi)(y) = \int dx\, \delta(x-f^{-1}(y)) \phi(x)
  = \sum_{x:\ f(x)=y} \phi(x)
\end{equation}
(for the time being we assume that $f$ is a 1-dimensional map).
A variety of applications\rf{AACI,CRR93}, some of which we shall turn to
below, requires use of generalized \evOper s weighted by
integer powers of the stability of the orbit
\beq
\Lop_{(k)}(y,x) =   f'(x)^{k-1} 
                  \delta ( x - f^{-1}(y))
\,.
\ee{weight_L}
Let $f^n(x)$ be the $n$th iterate of the map, and
\beq
    \Lambda_i \,:=\,  \prod_{j=0}^{n-1} f'(f^{j}(x_i))
\ee{jacob}
be the linear stability evaluated at the periodic point $x_i$,
given by the product over the $n$ periodic points belonging to a
given $n$-cycle.
We shall denote the  $n$th iterate raised to $m$th power by
$f^n(x)^m$, and the derivative of the $n$th iterate by
${f^n}'(x) = {d\over dx} f^n(x)$.
Traces of the powers of this \evOper\ are given by
\beq
\tr \Lop_{(k)}^n \;=\; \sum_{x_i\inFix{n}}
\frac{\Lambda_i^{k-1}}{|1-1/\Lambda_i|}
\,,
\ee{trace_L}
where we have assumed that all cycles have stability eigenvalues
$\Lambda_i \neq 1$ strictly bounded away from unity.
A cycle is called attracting, neutral or repelling if
$|\Lambda_i| < 1$, $|\Lambda_i| = 1$, $|\Lambda_i| > 1$, respectively.
If the map is repelling and all $|\Lambda_i| > 1$,
we can drop the absolute value brackets in \refeq{trace_L} and 
{\em define} the trace by
\beq
T_n(k) := 
        \sum_{x_i\inFix{n}} \frac{\Lambda_i^{k}}{(\Lambda_i-1)}
\,.
\label{tr_L}
\eeq
A corresponding ``determinant'' or ``Zeta function''
is related to the traces by 
\beq
F(z,k) = \det(1-z\Lop_{(k)}) =
  \exp\left(-\sum_{n=1}^\infty {z^n \over n} T_n(k) \right)
  = \sum_{n=0}^{\infty} a_n(k)z^n\,,
\ee{Fred_d}
and the spectrum of the \evOper\ is given by the zeros of $F(z,k)$.
In applications\rf{AACI,CRR93} determinants 
are often preferable to the trace sums.
For a lack of a better term we
shall refer to this function as the ``\Fd'' of the \evOper\ \refeq{weight_L} even
though, strictly speaking, the term should be used only for determinants of
operators proven to be trace class.

The main difference between the generalized \evOper s defined by
\refeq{weight_L},
and the operators that usually arise in the dynamical systems theory 
is that instead of the ``natural measure'' $1/|\Lambda_p|$ 
we deal here with a ``signed'' and possibly 
complex cycle weight  $1/\Lambda_p$.
There are several important applications of \evOper\ formalism
based on such ``signed'' measures, in
particular the periodic orbit formulas for
the spectrum of the period-doubling operator, \refsect{s_period-doub}.

From now on we {\em define} the trace of an 
\evOper\ $\Lop$ by
the above trace sum \refeq{tr_L} over 
{\em all cycles, real and complex}.
This choice of the cycle weight
is motivated by the contour integral sums
to which we turn now. 

\section{Contour integral evaluation of periodic orbits sums}
\label{s_Contour_int_sums}

A sum that runs over zeros of a function can be cast as
a contour integral residue calculation.
Periodic orbit formulas are sums over zeros of the 
periodicity condition $x-f^n (x)=0$,  
therefore we consider  contour integrals of type 
\[
T = \Cint{x} \frac{h(x)}{f(x)-x}  
\]
where $f$ and $h$ are polynomials, and the contour encloses
all zeros of $f(x)-x$.
By shrinking the contour to a sum of contours around 
individual zeros,
\reffig{f-CmplxContour}, and using the Cauchy formula
\[ \Cint{x} \frac{1}{x^k}=\delta_{k,1}
\]
we obtain a sum over fixed points $x_i$ of the map $f$:
%
\begin{figure}
\centerline{\epsfig{file=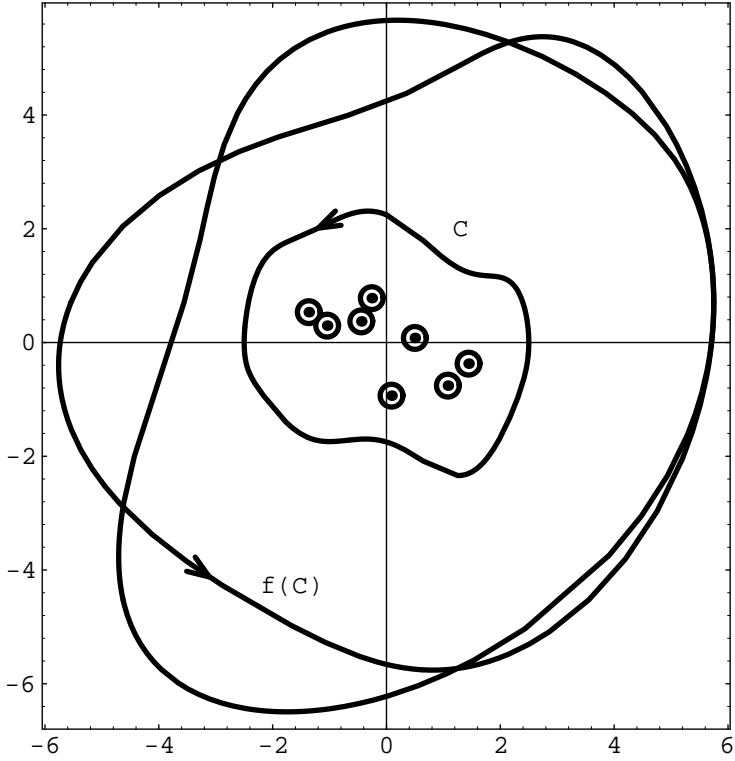,width=10cm}}
\caption[]{
Shrinking the contour $C$ to a sum of contours around
individual zeros leads to a sum over fixed points $x_i$ of the map $f$.
For a sufficiently large contour $C$, the iterate $C \to f(C)$ 
wraps twice around the periodic point set, leading to the key formula
\refeq{iter_contour}.
In this figure the map is $f(x) = x^2 -c$, $c = 0.5 - 0.7 i$, and
the 8 periodic points plotted are the zeros of $f^3(x)-x=0$. 
}
\label{f-CmplxContour}
\end{figure}
%
\beq
T = \sum_{x_i\inFix{}}
\frac{h(x_i)}{f'(x_i)-1}   
\,.
\label{contour0}
\eeq
Note that this cycle sum includes all periodic
points, real as well as complex. For $f^n$, $n\to\infty$ these points
fill out the Julia set, the closure of the set of 
the repelling periodic points\rf{deva87}.
We assume that all zeros are simple; multiple zeros can be treated
as well.
Here we shall compute
\beq
T{(k)} = \Cint{x} \frac{f'(x)^{k}}{f(x)-x} 
\ee{contour6}
for $k = 0,1,2,3,\dots$ (we turn to the $k=-1$ case in
\refsect{s_Julia_esc}).

The key observation is that
$T$ can be evaluated by pushing the contour to
$|x| \rightarrow \infty$, where the individual fixed points 
play no role.
When $k>0$, the following expansion of \refeq{contour6} is convenient:
\bea
  T{(k)} &=& \Cint{x} \frac{f'(x)^k}{x} \frac{x/f(x)}{1-x/f(x)}
                \continue
         &=&  \Cint{x} \frac{f'(x)^k}{x}
          \left(\frac{x}{f(x)}+\left(\frac{x}{f(x)}\right)^2+\cdots
          +\left(\frac{x}{f(x)}\right)^k\right)
\,.
\label{contour3}
\eea
For a polynomial this is an exact formula;
we use the fact that for sufficiently large $|x|$ and
order $N \geq 2$ of the polynomial $f$,
$(1-x/f(x))^{-1}$ can be expanded as a
geometric series.
Terms with powers of ${(x/f(x))^r}$ higher than $ r>k $ are of order at least
$O(x^{-N})$ and do not contribute  to 
the $|x| \rightarrow \infty$ contours.
Conversely, if $f$ is an analytic function, no such finite series
representation is immediately apparent.

By rescaling and translation any  polynomial of order 
$N$ can be brought to form
$f(x)=x^N - c\,x^{N-2} - \dots$ . Hence the last term is given by
\beq
\Cint{x} {1 \over x}
\;\frac{f'(x)^{k}}{(f(x)/x)^k}
 = \Cint{x} {1 \over x}
     \frac{N^k x^{k(N-1)}(1+O(x^{-2}))}{x^{k(N-1)}(1+O(x^{-2}))}
 = N^k \,.
\label{contour5}
\eeq
The first term in \refeq{contour3}  is a sum over the zeros of $f(x)$
\beq
C{(k)} :=
 \Cint{x} \frac{f'(x)^{k}}{f(x)}
 =  \sum_{f(x)=0} f'(x)^{k-1}
\,.
\ee{10a}
By the chain rule \refeq{jacob}, the $n$th iterate sum
is given by
\beq
 C_{n}(k) = \sum_{f^n(x)=0} {f^n}'(x)^{k-1}
 =  \sum_{f^{n-1}(y)=0} {f^{(n-1)}}'(y)^{k-1}
   \sum_{f(x)=y} {f}'(x)^{k-1}
\,.
\ee{10b}
For $k \leq 3$, the sum over zeros of $f(x)=0$ can be replaced\rf{Levin89} 
by the pre-images $f(x)=y$ of arbitrary constant $y$, as for $k=2,3$ the
difference
\bea
 &\Cint{x} \left\{ \frac{f'(x)^{k}}{f(x)-y} - \frac{f'(x)^{k}}{f(x)}
          \right\}
 &\continue
 &=
\sum_{j=2}^k y^{j-1}  \Cint{x}
     \frac{\left(N x^{(N-1)}-(N-2)c\,x^{N-3} - \dots\right)^k}
          {\left(x^{N}-c\,x^{N-2} - \dots\right)^j}
 &
\label{10c}
\eea
vanishes, and hence
by induction
the $n$th iterate sum \refeq{10b} is given by\rf{Levin89}
\beq
C_{n}(2) = C{(2)}^n 
\,\qquad
C_{n}(3) = C{(3)}^n 
\,.
\ee{12a}
The value of the sum $C{(k)}$ depends on the parametrization of the
particular polynomial.
For example, for the cubic polynomial 
\beq
f(x) = x^3 -c\,x -d
\ee{14a}
we have
\[
C{(2)} =  \Cint{x} {(3x^2-c)^2 \over x^3 -c\,x -d} =
   \Cint{x} 3^2x \left(1- {2c \over 3 x^2} + {c \over x^2} + \dots\right) 
         = 3c
\,.
\] 

If the polynomial is even, $f(x)=g(x^2)$, the
terms for $(x/f(x))^j$ for which $k+j$ is odd vanish, as 
\beq
   \Cint{x}\frac{f'(x)^k}{x} \left( \frac{x}{f(x)} \right)^j
   = 2^k\Cint{x\,x^{k+j-1}}\frac{g'(x^2)^k}{g(x^2)^j}
\,.
\ee{even-odd}
For example, for even polynomials $C{(2)}=0$. 
For general $k$ there are no formulas of the above type
for the other terms in the expansion \refeq{contour3}. 

\subsection{Sum rules for periodic orbits of any polynomial}

Our periodic orbits sum rules are
obtained by pushing the contour to
$|x| \rightarrow \infty$ in \refeq{contour6}.
The first periodic orbits sum rule  for 1-$d$ maps follows from
\[
T{(0)} = \Cint{x} \frac{1}{f(x)-x}
 = \Cint{x} \frac{1}{x^{N}(1+O(x^{-1}))} =0
\,,
\]
hence
\beq
\sum_{i\inFix{n}} \frac{1}{\Lambda_i-1} = 0
\,.
\label{tr_0}
\eeq
The second and third periodic orbits sum rules follow from \refeq{contour5}:
\bea
T_n{(1)} &=& \Cint{x} \frac{{f^n}'(x)}{f^n(x)-x}
 = \sum_{i\inFix{n}} \frac{\Lambda_i}{\Lambda_i-1}
= N^n
\label{1st_sr}\\
T_n{(2)} &=& \Cint{x} \frac{{f^n}'(x)^2}{f^n(x)-x} =
 \sum_{i\inFix{n}} \frac{\Lambda_i^2}{\Lambda_i-1} =  N^{2n}
        + C{(2)}^n
\,.
\label{2nd_sr}
\eea
The corresponding \Fd s follow from \refeq{Fred_d}:
\bea
F(z,0) &=& 1, \qquad
F(z,1) =1-zN, \continue
F(z,2) &=& (1-zN^2)(1-zC{(2)})
\,.
\label{Fd:0:1:2}
\eea
The $T_n{(0)}$, $T_n{(1)}$ sum rules will be generalized to
$d$-dimensions in \refsect{s_Mult-dim}.
Using \refeq{contour3} we obtain the fourth sum rule 
\beq
T_n{(3)} =
C{(3)}^n \; +\; \Cint{x}\frac{x {f^n}'(x)^3} {f^n(x)^2} \; + \; N^{3n}
 = 
\sum_{i\inFix{}} \frac{\Lambda_i^3}{\Lambda_i-1} 
\,.
\label{3rd_sr}
\eeq
For even polynomials the second term vanishes.

These are exceedingly simple sum rules and expressions for
\Fd s, and we can already 
discern both their utility and the ways in which they might
fall short of what we need in order to perform efficient dynamical
systems computations. By taking into account all periodic points,
not just the real ones, these sum rules dispense with the infinite
intricacies of controlling the parameter dependence of
symbolic dynamics of systems as simple as a parabola.
However, there are clearly very important
physical effects that our sum rules are blind to: the most striking is
the role played by bifurcations. While at a bifurcation real dynamics
changes qualitatively, going from hyperbolic through intermittent to
a stable attractor, the corresponding complexified 
``signed'' sum rules are arranged precisely in
such a way that these effects cancel.

Starting with $T{(3)}$, the sum rules cannot be cast in form applicable to
arbitrary polynomials, so we now specialize to the quadratic polynomials.
We refer the reader interested in the 
\evOper s weighted by more general rational polynomials 
to \refref{Levin92}.

\section{Sum rules for the quadratic map}
\label{s_tr_3}

The quadratic polynomial
\beq
f(x) = x^2 - c
\ee{fateau}
is the simplest example of a non-linear 1-$d$ 
mapping
(note the unconventional sign of the parameter $c$, chosen to
simplify our final formulas).
By the chain rule \refeq{jacob}
for the derivative of an iterated function, the
stability of an n-cycle $\{x_1, x_2, \cdots, x_n\}$ is given by
\beq
\Lambda = 2^n x_1 x_2\cdots x_n
\,.
\ee{Chain_r}

We shall now derive an explicit formula for the sum rules and \Fd s 
for transfer operators 
\refeq{weight_L}
weighted by integer powers of the unstable
eigenvalue $\Lambda^k$  in
\refeq{tr_L}. The derivation is based on the following
simple observation
which relates contour integrals of functions of the $n$th iterate $f^n(x)$ to 
contour integrals of functions of $f^{n+1}(x)$:
for a quadratic polynomial and a sufficiently large contour, 
a change of the integration variable $x \to y =f(x)$
wraps the contour around twice
\beq
  \oint_\gamma \! \frac{dx\,f'(x)}{2\pi i} h(f(x)) =
  2 \oint_\gamma \! \frac{dy}{2\pi i} \, h(y).
\ee{iter_contour}
The factor 2 arises because for large $|x|$ the phase of $f(x) \approx x^2$
advances at twice the speed of the phase of $x$, see 
\reffig{f-CmplxContour}.
Reconsider now the sum \refeq{contour3} with the last term evaluated
as in \refeq{contour5}:
\[
  T_n{(k)} = \Cint{x} \frac{{f^n}'(x)^{k}}{x}
          \sum_{s=1}^{k-1} \left(\frac{x}{f^n(x)}\right)^{s} + 2^{k n}
\,.
\]
The individual terms in the sum
are the diagonal terms of the $[(k-1) \times (k-1)]$ matrix
\beq
  {\bf A}^{(n)}_{rs} := 
     \Cint{x} \frac{x^{s-1} }{f^n(x)^{r}} {f^n}'(x)^{k}
\,, \qquad r,s = 1,2,\cdots,k-1 \,,
\ee{matrixA}
so the $n$th iterate trace can be expressed as
\[
T_n{(k)} = \tr {\bf A}^{(n)} + 2^{k n} 
\,.
\]
Trivially, for even polynomials $A_{rs} \neq 0$ if
$k$ and $s-1$ are either both odd or both even.
The motivation for introducing the matrix ${\bf A}^{(n)}$ 
is the observation that its elements are a convenient basis for relating 
the successive trace sums $T_n{(k)} \to T_{n+1}{(k)}$.
By \refeq{even-odd} 
\[
  {\bf A}^{(n+1)}_{rs}
    = \Cint{x f'(x)}\frac{x^{s-1} f'(x)^{k-1}}{f^n(f(x))^{r}} 
                     {f^n}'(f(x))^{k}       
\]
vanishes for even polynomial $f$ unless $s+k=$~even.
Hence we can substitute 
$x^{s-1}f'(x)^{k-1} = 2^{k-1} x^{s+k-2} = 2^{k-1} (y+c)^{s+k-2 \over 2}$, 
and applying \refeq{iter_contour} obtain 
\[
  {\bf A}^{(n+1)}_{rs}
    = 2^{k} \Cint{y}
      \frac{(y+c)^{k+s-2\over 2}}{f^{n}(y)^{r}}
                {f^{n}}'(y)^{k}     
\,.
\]
We note that this step relies on the very simple form of the
quadratic polynomial, so a generalization to
arbitrary polynomial mappings is not immediate.
Now expand binomially the $(y+c)$ term 
\[
{\bf A}^{(n+1)}_{rs} =
      2^{k} \sum_{p=0}^{k+s-2\over 2} {{k+s-2\over 2} \choose p} 
      c^{{k+s-2\over 2}-p} {\bf A}^{(n)}_{r,p+1} 
\]
and observe that the successive ${\bf A}^{(n)}$ 
are obtained by multiplication by 
the $[(k-1) \times (k-1)]$ dimensional matrix
\beq
{\bf L}_{(k)\, {rs}} := \cases{ \qquad 0 & if $k,s$ differ in parity 
                          \\
\bs
      {{k+s-2\over 2} \choose r-1} c^{{k+s\over 2}-r}
                             & otherwise.
                          }
\ee{B_ulige}
As ${\bf {\bf A}}^{(0)} = {\bf 1}$, the $n$th level trace sums are
given by the trace of 
$
  {\bf {\bf A}}^{(n)} = 2^{k n}{\bf L}^n_{(k)}    
$,
and the \Fd\ \refeq{Fred_d} of the \evOper\ \refeq{weight_L} is given by
\beq
F(z/2^{k},k) = (1-z)\det\left(1-z {\bf L}_{(k)}\right) 
\ee{Fred_B}
The first 10 such determinants are tabulated in \reftab{t_quadr_map}, and an
example is given in \ref{s_Tr_L_7}.
This is the main result of this section: we have 
obtained an explicit formula for the
spectrum of the  \evOper\ $\Lop_{(k)}$ for any positive integer
$k$ in terms of the eigenvalues of a {\em finite} matrix ${\bf L}_{(k)}$. 
As by the symmetry of $f$ 
the entries in the half of the columns of \refeq{B_ulige} 
vanish, in computations it is convenient to distinguish
the odd and even cases, and take ${\bf L}_{(k)}$ to be a $[l \times l]$
matrix, $i,j =0,1,2,\cdots,l-1$ :
\beq
{\bf L}_{(k)\,ij} := \cases{ 
       {l+j \choose 2i} c^{l+j-2i} 
                                & odd~ $k=2 l+1$
                           \\
\bs
       {l+j+1 \choose 2 i+1 } c^{l+j -2i}
                                & even $k=2 l+2$
\,.
                          }
\ee{B_lige}
These finite matrices were first introduced by 
Levin,  Sodin, and  Yuditskii, section 4 of \refref{Levin92}.
%
\begin{table}
\caption[]{
\Fd s \refeq{Fred_B} for the quadratic map \evOper s 
 $\Lop_{(k)}$, for $k$ a positive integer.
        }
\vskip 1pt
\begin{center}
{\small
\begin{indented}\item[]
\begin{tabular}{@{}rl}
\br
$ k$  & ~~~~~~~~~~~~~~~~ $\det\left(1-z {\bf L}_{(k)}\right)$\\
\hline
2  & $ 1 $ \\
3  & $ 1-c\,z $ \\
4  & $ 1-2\,c\,z $ \\
5  & $ 1-(3+c)\,c\,z+2\,{c^3}\,{z^2} $\\
6  & $ 1-(4+3\,c)\,c\,z+8\,{c^3}\,{z^2} $\\
7  & $
 1 - (5 + c)(1 + c)\,c \,z +
    \left( 20 + 5\,c + 3\,{c^2} \right)\,{c^3} \,{z^2} - 8\,{c^6}\,{z^3}
                    $ \\
8  & $
 1 - \left( 6 + 10\,{c} + 4\,{c^2} \right) \,c \,z + 
     \left( 40 + 24\,{c} + 20\,{c^2} \right) \,{c^3}\,{z^2} - 
     64\,{c^6}\,{z^3}
                    $ \\
9  & $
1 - \left( 7 + 15\,{c} + 10\,{c^2} + {c^3} \right) \,c\,z + 
    \left( 70 + 70\,{c} + 82\,{c^2} + 14\,{c^3} + 4\,{c^4}
          \right) \,{c^3} \,{z^2} 
                    $ \\ 
   & $
       ~~~~~~~~~~~~~~~~~~~~~~ ~~~~~~~~~~~~~~~~~~~~~~
        - \left( 280 + 64\,{c} + 28\,{c^2} + 
       20\,{c^3} \right) \,{c^6}\,{z^3} + 64\,{c^{10}}\,{z^4}
                    $ \\
10  & $
1 - \left(8 + 21\,{c} + 20\,{c^2} + 5\,{c^3} \right)\,c \,z + 
\left( 112 + 160\,{c} + 250\,{c^2} + 98\,{c^3} + 40\,{c^4} 
        \right) \,{c^3}\,{z^2}  $ \\
   & $
       ~~~~~~~~~~~~~~~~~~~~~~ ~~~~~~~~~~~~
       - \left(896 + 512\,{c} + 320\,{c^2} + 
       280\,{c^3} \right)\,{c^6}\,{z^3} + 1024\,{c^{10}}\,{z^4}
                    $ \\
$\cdots$ & $\cdots$\\
\br
\end{tabular}
\end{indented}
}  
\end{center}
\label{t_quadr_map}
\end{table}

\subsection{Dynamical zeta functions}

Using a different approach,
Hatjispiros and Vivaldi\rf{HV93} 
have introduced a family of dynamical zeta functions 
for complex quadratic polynomials, 
and conjectured 
that these zeta functions are of a  particularly simple rational form.
We note that the formula \refeq{Fred_B} proves the conjectured rationality
of this family of \dzeta s.
However, their method of evaluating the determinants is still of
considerable interest, as it appears to be more efficient for evaluation
of $\det\left(1-z {\bf L}_{(k)}\right)$ for high $k$ than the
direct evaluation undertaken here.

For purposes of comparison with \refref{HV93} we need to 
relate \dzeta s to the \Fd s given above.
For 
$|\Lambda| >1$,
we expand the weight in \refeq{tr_L}
\[
{ 1 \over \Lambda^r (1-1/\Lambda^{r}) }
 =   {1 \over \Lambda^r} \sum_{j=0}^\infty \Lambda^{-jr}
\,\,  ,
\]
and obtain the product representation of the \Fd\ for 1-$d$ maps
\[
F{(z,k)} = \exp\left( - \sum_p \sum_{r=1}^\infty {z^{n_p r} \over r}
                  { 1 \over \Lambda^{r k}_p (\Lambda^r_p-1) }
        \right)
=  \prod_p\prod_{j=0}^{\infty} \left(1-{z^{n_p}\over \Lambda_p^{k+j+1}}\right)
\,\,,
\] 
where the $p$ product goes over all prime (not self-retracing) cycles.
The \dzeta\ is defined\rf{ruelle} as
\beq
1/\zeta_{(k)}=\prod_p \left(1-{z^{n_p}\over \Lambda_p^{k+1}}\right)
\,,
\ee{wdynzeta}
so for 1-$d$ mappings it can be expressed as a ratio of two \Fd s
\beq
1/\zeta_{(k)} = { F{(z,k)} / F{(z,k+1)}}
 \,\, .
\ee{zeta_k}
The zeta functions of \refref{HV93}
are indexed and normalized differently;
here we follow the conventions used in \refref{CRR93}.
Our \Fd s are related to the polynomials $G$ of  \refref{HV93} by 
$\det\left(1-z {\bf L}_{(k)}\right)=G_{k-1}(z/2,-c)$, and our 
explicit formula \refeq{Fred_B} is a proof of their conjectures.

\section{Julia set escape rate}
\label{s_Julia_esc}

So far we have obtained explicit finite expressions for the
spectral determinants of signed transfer operators
$\Lop_{(k)}$ with weight $(f')^k, k\geq 0$. These results can
easily be generalized to arbitrary polynomial weights
.
However, the methods outlined in \refsect{s_Contour_int_sums}
fail for weights $(f')^k$ with $k<0$. Especially the case $k=-1$
is physically interesting, because the largest
eigenvalue of the operator gives the escape rate
$\gamma$ from a given enclosure
$\Gamma$, which is defined by the fraction of initial
points that stay in $\Gamma$ after $n$ iterations
\rf{cycl_book}
\begin{equation}
  \label{escape}
  e^{-n\gamma} \approx
  \frac{\int_{\Gamma} dx\, dy\, \delta(x-f^n(y))}{\int_{\Gamma}dx}.
\end{equation}
The signed version of this transfer operator is defined
with the weight $1 \over {f'(y)}^2$ and by summing over
all real and complex pre-images of $x$:
\begin{equation}
  \label{k=-1.2}
  (\Lop_{(-1)} \phi)(x) = \sum_{y:f(y)=x}
  \frac{\phi(y)}{f'(y)^2},\quad x, y\in {\bf C}\, .
\end{equation}
If the Julia set is real the two operators coincide.

The spectral determinant
of \refeq{k=-1.2} has been computed in \refref{Levin92} for quadratic
maps $f(x)=x^2-c$. If $c$ lies outside the Mandelbrot set
$M$, the critical point is attracted by infinity and
the spectral determinant is given by the entire
function
\begin{equation}
  \label{specdet}
  F(z, -1) = 1 + \sum_{n=1}^{\infty}
  \frac{z^n}{2^nf(0)f^2(0)\cdots f^n(0)},
\end{equation}
which involves only forward iterates of the critical point
of $f$. If $c$ lies inside the Mandelbrot set, the 
critical point is attracted by a limit cycle. If the 
stability of this cycle is different from zero,
\refeq{specdet} is the form of the spectral determinant
for small $z$ and can be meromorphically continued.
If on the other hand the critical point is periodic with
period $n_p$, the spectral determinant is a polynomial \rf{Levin92}.

In this section first we outline the derivation of this result (for details
consult \refrefs{Levin92}). Then we generalize it and show that
it is correct also for a dense set of maps on the boundary of the
Mandelbrot set.

The Julia set $J$ is the closure of all repelling periodic orbits.
If $c$ lies outside the Mandelbrot set $M$, the Julia set resembles
the Cantor dust. To prove \refeq{specdet} we have to use 
analyticity properties of the transfer operator and its adjoint.
One starts by defining the evolution operator $\Lop_{(-1)}$ on the
space of functions $\psi$ which are locally analytic in a small
neighbourhood around
$J$. Then one can prove that the spectrum of $\Lop_{(-1)}$ is 
a point spectrum with the only possible condensation at zero.
The dual or adjoint operator $\Lop_{(-1)}^{\star}$ of the
transfer operator $\Lop_{(-1)}$, defined on densities $\phi^{\star}$ 
dual to
the functions $\psi$, then has the 
same spectrum. The duality is with respect to the bilinear form
\begin{equation}
  \label{dual}
  \phi^{\star}(\psi) = \oint_{\gamma} \frac{dx}{2\pi i} 
  \phi(x) \psi(x)\, ,\,\mbox{and}\ \ \phi(x) = \phi^{\star}(\frac{1}{x-.})
\end{equation}
which is natural if one deals with analytic functions. The 
{\it function} $\phi(x)$ is holomorphic on the complement of $J$ and
vanishes at infinity. The contour $\gamma$ encircles the Julia 
set and lies in the common domain of analyticity of $\phi$ and $\psi$.
The spectral equation $\phi^{\star}-z\Lop_{(-1)}^{\star}\phi^{\star}=0$
can then be written as
\begin{equation}
  \label{speceq}
   \oint_{\gamma} \frac{dx}{2\pi i} \psi(x)
  \left( \phi(x) - z\frac{1}{f'(x)}\phi(f(x))\right) = 0, 
\end{equation}
for all $\psi$.
Therefore
\begin{equation}
  \label{hol}
  R(x) := \phi(x) - z\frac{1}{f'(x)}\phi(f(x))
\end{equation}
has to be holomorphic around $J$ and can have poles only
if $\frac{1}{f'(x)}$ has poles, because $\phi$ is analytic
outside $J$.

Specializing to the map $f(x) = x^2-c$, the adjoint
equation \refeq{speceq} leads to 
\begin{equation}
  \label{holquad}
  R(x) = \phi(x) - \frac{z}{2x}\phi(x^2-c).
\end{equation}
We conclude that $R$ vanishes at infinity because $\phi$
does. Furthermore $R$ can only have a pole of order one
at zero, the unique critical point $x_c$ of $f$.
That means we have either $R(x)=\frac{1}{x}$
(up to a multiplicative constant) or $R(x)=0$.
$R(x)=0$ is contradictory
because it yields eigenfunctions
equal to zero. By iterating \refeq{holquad} after solving it
for $\phi(x)$ we get
\begin{equation}
  \label{eigfunc}
  \phi(x) = {1\over x} + \sum_{n=1}^{\infty}
  \frac{z^n}{2^nf^n(x)\prod_{j=0}^{n-1}f^j(x)}.
\end{equation}
Now note that $\phi$ is only an eigenfunction
if it is holomorphic at the poles of $R$, that means
at zero in our case ($f'(x_c)=0$). Therefore
\begin{eqnarray}
  \label{res}
  0 &=& \mbox{Res}_{x=0}[\phi(x)] \nonumber\\
  &=& 1 + \sum_{n=1}^{\infty} \frac{z^n}{2^nf(0)f^2(0)\cdots f^n(0)} = F(z, -1)
\end{eqnarray}
is an equation for the characteristic values of $\Lop_{(-1)}$ meaning
that \refeq{eigfunc} is an eigenfunction if and only if $z$ is an
eigenvalue. For a generalization of this method to arbitrary
rational maps see \rf{Levin92}.

The form \refeq{specdet} for the spectral determinant for
the map $f(x)=x^2-c$ holds true, if $c$ lies outside
or inside the Mandelbrot set $M$ \rf{Levin92}.
However,
the method fails
if $c$ lies on the border of $M$, and in this case the form of the spectral
determinant is unknown.

As a generalization to \rf{Levin92} we show here 
that on a dense subset
of the border of $M$, the so called Misiurewicz points,
formula 
\refeq{specdet}
still applies.
The Misiurewicz points are defined by the requirement that
\begin{figure}
\centerline{\epsfig{file=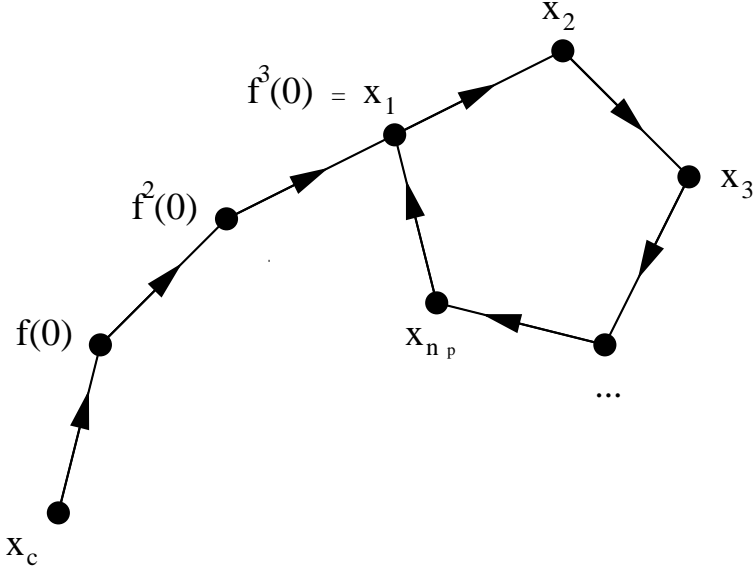,width=10cm}}
\caption[]{The critical point $x_c=0$ is a preperiodic point
  with transient time $t=3$ and period $n_p=5$. In this case
  the parameter $c$ of the quadratic map is called a
  Misiurewicz point. Note that the periodic orbit is a
  repeller.}
\label{prep}
\end{figure}
the critical point $x_c=0$ is preperiodic \rf{PJS}. In
\reffig{prep} we introduce the notation for preperiodic points.
If we take the validity of \refeq{specdet} for granted,
a straightforward calculation yields that 
for Misiurewicz points the spectral determinant is in fact 
a rational function composed of a transient part and a periodic
contribution
\begin{eqnarray}
  \label{misi}
  F(z, -1) &=& 1 + \sum_{n=1}^{t-2}
  \frac{z^n}{2^n\prod_{j=1}^n f^j(0)}
  +\frac{z^{t-1}}{2^{t-1}\prod_{j=1}^{t-1} f^j(0)}
  \frac{\sigma(z)}{\left(1-\frac{z^{n_p}}{\Lambda_p}\right)}\, .
\end{eqnarray}
Here
\begin{equation}
  \label{sigma}
  \sigma(z) = 1 + \frac{z}{2x_1}+
  \frac{z^2}{4x_1x_2} + \ldots +
  \frac{z^{{n_p}-1}}{2^{{n_p}-1}x_1\cdots x_{{n_p}-1}}\, ,
\end{equation}
$\Lambda_p=2^{n_p}x_1\cdots x_{n_p}$ is the stability of the periodic orbit,
${n_p}$ is the period and $t$ the transient time of the Misiurewicz
point.
The $n_p+t-2$ zeroes of $F(z, -1)$ are the characteristic values
of $\Lop_{(-1)}$. $F$ has $n_p$ poles which lie on a cycle around
the origin in the complex $z$-plane. As the period $n_p$ of the orbit
increases to infinity, these poles built up a wall which limits
the region of analyticity.

Presumably, formula \refeq{misi} can be proved by a generalization
of the arguments in \rf{Levin92}.
Here we check it using cycle expansion, i.e. the power series
representation \refeq{Fred_d} of the \Fd\rf{AACI,cycl_book}.
As the first example we compute the spectral determinant for the Ulam map
$f(x) = x^2 - 2$. For this map the critical point $x_c = 0$ iterates via
$f(0)=-2$ to the unstable fixed point $x^{\star}=2$. That means we have
$t=2$ and $n_p=1$, and formula \refeq{misi} yields
\begin{equation}
  \label{ulam}
  F(z,-1) = 1  -\frac{z}{4} \frac{1}{1-z/4} = \frac{1-z/2}{1-z/4}
\end{equation}
in agreement with the cycle expansion result of \refref{AACI},
where this formula is derived by observing
that for all cycles the stability
is $\vert\Lambda_p\vert=2^{n_p}$, with exception of the fixed point
$x_0$ for which $\Lambda_0=4$. However, the result for a general
Misiurewicz polynomial is not as trivial.

Next we check \refeq{misi} by a numerical computation of the
spectral determinant
via complex cycle expansion. The method is described in detail in \rf{R97}.
Here we will only state the results. As a first example we study the case
$c=1.5436890126920763616\ldots$, for which the critical point $x_c=0$ iterates after
three steps into a fixed point $x^{\star}=c(1-c)$, so $t=3$ and $n_p=1$.
Therefore we expect a spectral determinant of the form
\begin{equation}
  \label{quad}
  F(z, -1) = 1 + \frac{z}{2f(0)}
  + \frac{z^2}{4f(0)f^2(0)}
  \frac{1}{\left(1-\frac{z}{2x^{\star}}\right)}
\end{equation}
with $f(0)=-c$ and $f^2(0) = -x^{\star} = c(c-1)$.
In \reftab{tab} the coefficients $a_n$ of the cycle expansion for the spectral
\begin{table}[t]
  \small
  \begin{center}
    \begin{tabular}[c]{|r|c|c|} \hline
      \multicolumn{1}{|c|}{$n$} & \multicolumn{1}{c|}{$a_n$} &
      \multicolumn{1}{c|}{$\frac{a_n}{a_{n+1}}$} \\ \hline
      0 &1.00000000000000000  & 3.087378025384153\\
      1 &-0.3238994356305212 & 1.678573510428349\\
      2 &-0.1929611265865065 &-1.678573510428322\\
      3 & 0.1149554222008833  &-1.678573510428341\\
      4 &-0.0684839963735331 &-1.678573510428292\\
      5 & 0.0407989259618778  &-1.678573510428285\\
      6 &-0.0243057129809394 &-1.678573510428318\\
      7 & 0.0144799812638157  &-1.678573510428306\\
      8 &-0.0086263611178524 &\\ \hline
    \end{tabular}
    \caption{Spectral determinant for a
      chaotic quadratic map up to order
      8 in complex cycle expansion. For $n>1$, the ratio 
      $a_n/a_{n+1}$ of the cycle expansion coefficients
      \refeq{Fred_d} computed from $2^n$ periodic points
      equals $\Lambda=2c(1-c)=-1.6785735104283$ to machine
      precision, in agreement with \refeq{quad} }
    \label{tab}
  \end{center}
\end{table}
determinant $F(z,-1) = \sum_{n=0}^{\infty} a_nz^n$ are depicted.
The second column of the table demonstrates that they indeed
decrease geometrically with the correct ratio. The remaining
coefficients can be checked numerically or
computed analytically from the traces of $\Lop_{(-1)}$ and $\Lop_{(-1)}^2$.
In either case we arrive at \refeq{quad}. Because the numerator 
of the spectral determinant is quadratic there are two 
characteristic values, namely
$z=c^2*(1-c) = -1.295597742522084$ and $z=2$. That means the
spectrum is real and there is a real pole at 
$z=2c(1-c) = -1.678573510428321$.
The escape rate \refeq{escape} is
given by the trivial largest eigenvalue
$\gamma=\log(2)$.

As our last example we take the map $f(x) = x^2+i$, for which
$x_c=0$ has the trajectory
\begin{eqnarray*}
  0\mapsto i\mapsto x_1=-1+i\mapsto x_2=-i\mapsto -1+i\mapsto \ldots,
\end{eqnarray*}
so $t=2$ and $n_p=2$. The numerical cycle expansion\rf{R97} yields
\begin{equation}
  \label{mpi}
  F(z, -1) = \frac{\frac{z}{2i}-
    \frac{z^2}{4+4i}}{1-\frac{z^2}{4+4i}}\, ,
\end{equation}
where by \refeq{misi} we had expected
\begin{equation}
  \label{expect}
  F(z, -1) = 1 + \frac{z}{2f(0)}\frac{1+
    \frac{z}{2x_1}}{1-\frac{z^2}{4x_1x_2}}\, .
\end{equation}
After short inspection the two expressions turn out to be the same.

Any number of such examples can be worked out in the same way.
In every case a scaling of the numerical coefficients
confirms the rational form of the Fredholm determinant for
the Misiurewicz polynomials.

\section{Multi-dimensional polynomial mappings}
\label{s_Mult-dim}

In the multi-dimensional case a typical contour integral representation
of a periodic orbit sum is of form
\beq
T{(0)}=\Cint{x_1} \Cint{x_2} \cdots \Cint{x_d} 
\prod_{\alpha=1}^d \frac{1}{ f_\alpha(x) - x_\alpha}
\,.
\ee{t_0_mult}
For example, in 2 dimensions the 
contour integral picks up a contribution from
each fixed point $i$
\[f_1(x_i,y_i)-x_i=0\,,\quad f_2(x_i,y_i)-y_i=0
\,.
\]
The integral can be
converted into a sum of local contour integrals around 
linearized neighbourhoods of the fixed points
$(x_i,y_i)$ 
\bea
T{(0)}&=&\sum_i\Cint{z_{i1}}\Cint{z_{i2}}
\frac{1}{(z_{i1}-J_{11}z_{i1}-J_{12}z_{i2})(z_{i2}-J_{21}z_{i1}-
J_{22}z_{i2})}
\,,
\nnu
\eea
where $x=x_i+z_{i1}$,  $y=y_i+z_{i2}$, and ${\bf J}_i$ is the 
Jacobian matrix
\[
\MatrixII{ J_{11}}{J_{12}}{J_{21}}{J_{22}} =
       \MatrixII{ \partial_x f_1(x,y)}{ \partial_y f_1(x,y)}
                { \partial_x f_2(x,y)}{ \partial_y f_2(x,y)}
                                                       _{(x,y)=(x_i,y_i)}
\,.
\]
Completing the integrals we get for 2 dimensions
\[
T{(0)}=\sum_{i\inFix{}}
\frac{1}{(1-J_{11})(1-J_{22})-J_{12}J_{21}}
\,,
\]
and in $d$ dimensions
\beq
T{(0)}=\sum_{i\inFix{}} \frac{1}{\det(1-{\bf J}_i)} = 0
\,,
\ee{tr_0_d}
where ${\bf J}_i$ is the Jacobian (monodromy matrix) of the $i$th
fixed point.
Pushing all contours
in  \refeq{t_0_mult} off to infinity yields the sum rule $T{(0)}= 0$.
Just as in the 1-dimensional case \refeq{tr_0},
$T{(0)}$ vanishes when all fixed points, real
and complex, are included in the sum.

\subsection{A sum rule related to the fast kinematic dynamo}
\label{s_kin_dyn}

Our second multi-dimensional 
sum rule arises in the study of a 2-$d$ Poincar\'e map
of a model of the fast kinematic dynamo\rf{dynamo}. 
The fast kinematic dynamo is a problem of passive vector field advection:
the dynamo rate is determined by the overall growth of a small vector field
embedded into the flow.
In this case the \evOper\ \refeq{weight_L} is
weighted by the Jacobian matrix ${\bf J}^t(x,y)$, a multiplicative
function evaluated along the trajectory with initial point $(x,y)$.
For 2-$d$ maps this \evOper\ leads to traces of form 
$\tr \Lop = \sum \tr {\bf J}/|\det(1-{\bf J})|$. We can derive a sum rule
for the corresponding signed measure trace sum:
\bea
T{(1)} &=& \Cint{x}\Cint{y}
\frac{\tr {\bf J}(x,y)}{(x-f(x,y))(y-g(x,y))}
        \continue
&=& \sum_{i\inFix{}} \frac{\tr {\bf J}_i}{\det(1-{\bf J}_i)}
\,.
\nnu
\eea
The 2-$d$ matrix identity
\bea
\frac{\tr {\bf J}}{\det(1-{\bf J})} &=&
\frac{\tr {\bf J}}{1 -\tr {\bf J} +\det{\bf J}} 
   \continue &=&
\frac{1}{\det(1-{\bf J})} - \frac{1}{\det(1-{\bf J}^{-1})} -1
\nnu
\eea
yields our sum rule
\beq
T{(1)}= 
\sum_{i\inFix{}} \frac{\tr {\bf J}_i}{\det(1-{\bf J}_i)} =
         -N
\,,
\ee{tr_1_d}
where we have used the fact that $T{(0)}$ vanishes for both
the forward and the time reversed flow. 
As in the 1-$d$ case \refeq{1st_sr}, $T{(1)}$ counts the number of the fixed
points of the map; in polynomial maps this is given by the
order of the polynomial. 

\subsection{An application: numerical checks of cycle sums}

As an example for the  utility of sum rules for higher dimensional 
polynomial mappings, consider the 2-dimensional H\'enon map
\beq
x_{k+1} = 1-ax^2_k+y_{k}\,, \quad
y_{k+1} = b x_{k}
\,.
\ee{henon_m}
For the complete repeller case (all binary sequences are realized),
the H\'enon map is a realization
of the complete Smale horseshoe.
Cycle stabilities are easily computed numerically\rf{CRR93}.
We have verified numerically the sum rules 
\refeq{tr_0_d} and \refeq{tr_1_d} for a volume
preserving H\'enon map repeller ($b=-1$) by
substituting the cycle stabilities of cycles up to length 12 into
\bea
\sum_{i\inFix{n}} 
   \frac{1}{\Lambda_i(1-1/\Lambda_i)^2} 
&=& 0  
\label{1st-2dHam}\\
 \sum_{i\inFix{n}}
   \frac{\Lambda_i+1/\Lambda_i}{\Lambda_i(1-1/\Lambda_i)^2}
&=& 2^n 
\label{2nd-2dHam}
\eea
(here $\Lambda_i$ is the expanding eigenvalue of ${\bf J}_i$). 
The first sum can be used to check the accuracy of the
periodic orbit data used to compute the escape rate $\gamma$
of the H\'enon repeller
\[
   \sum_{i\inFix{n}}
   \frac{1}{|\Lambda_i| (1-1/\Lambda_i)^2} = e^{-n\gamma}
        + \mbox{~(non-leading eigenvalues)}
\,.
\]
If we add \refeq{1st-2dHam} to the above sum, for complete horseshoe all inverse
hyperbolic cycle contributions with
$\Lambda_i<0$ cancel, and hence only a half of the cycles suffices to
to compute the escape rate.
The second sum rule can be used 
to check the accuracy of the data used to estimate
the kinematic dynamo rate $\eta$ 
\[
   \sum_{i\inFix{n}}
   \frac{\Lambda_i+1/\Lambda_i}{|\Lambda_i| (1-1/\Lambda_i)^2} = e^{n\eta}
        + \mbox{~(non-leading eigenvalues)}
\,.
\]
It was a numerical discovery of these sum rules
computed in connection with the kinematic dynamo 
model of \refref{dynamo} that lead
us to the contour integral formulation and derivation of all 
other sum rules given in this paper.

\section{Farey map (or spin-chain) thermodynamics}
\label{s_CIRC}

\begin{figure}
\centerline{\epsfig{file=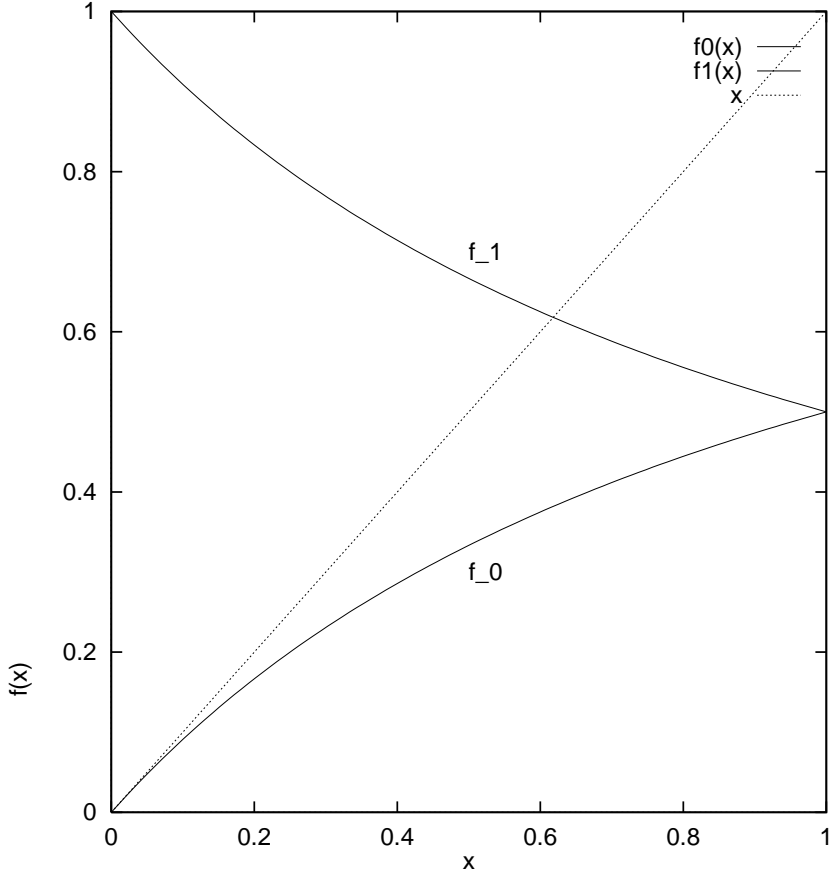,width=12cm}}
\caption[]{
The two branches of the inverse to the Farey map \refeq{far_present}. 
}
\label{f_farey_inv}
\end{figure}

So far, our examples were based on the contour integral
technique. In this section we shall use a very different,
number theoretic approach to compute the spectrum
of {\evOper}s. 
We shall consider the
Farey map, \reffig{f_farey_inv}, a combination of two 
Moebius transformations
%
%
\beq
f(x) = 
\left\{
\begin{array}{lcl}
x/(1-x) \qquad x &\in& [0,1/2)  \\
(1-x)/x \qquad x &\in& [1/2,1]  
\end{array}
\right.
\,\, .
\label{far_present}
\eeq
The Farey map arises in a variety of contexts: dynamics of circle maps,
dynamical renormalization theory, statistical mechanics, number theory.
For dynamical systems motivation we refer the reader to the review\rf{C92a}.
In particular, the Farey map plays the same role {\em vis-a-vis}
the shift circle map that the highly nontrivial circle map
presentation function\rf{feignonlin,CGV} plays in the golden-mean
renormalization theory, and provides a controllable setting 
to test some of the ideas that arise in that context.
Another motivation comes from statistical
mechanics, where Farey model thermodynamics corresponds to
a number-theoretical model of an infinite ferromagnetic spin chain
with effective $r^{-\alpha}$ interaction\rf{CK95}, with phase transition of
Thouless type at $\alpha=2$. For the purpose at hand the
Farey map is of interest to us because
\begin{itemize}
\item[(i)]
        again an infinity of sum rules can be obtained, but this time without the
        contour integration approach employed above
\item[(ii)]
        the sum rules apply to the positive measure $|\Lambda|^\tau$,
        in contradistinction to the signed measures of the previous examples
\item[(iii)]
        the theory can be extended to no-integer exponents $\tau$, with
        infinite range recursion relations.
\end{itemize}

\begin{figure}
\centerline{\epsfig{file=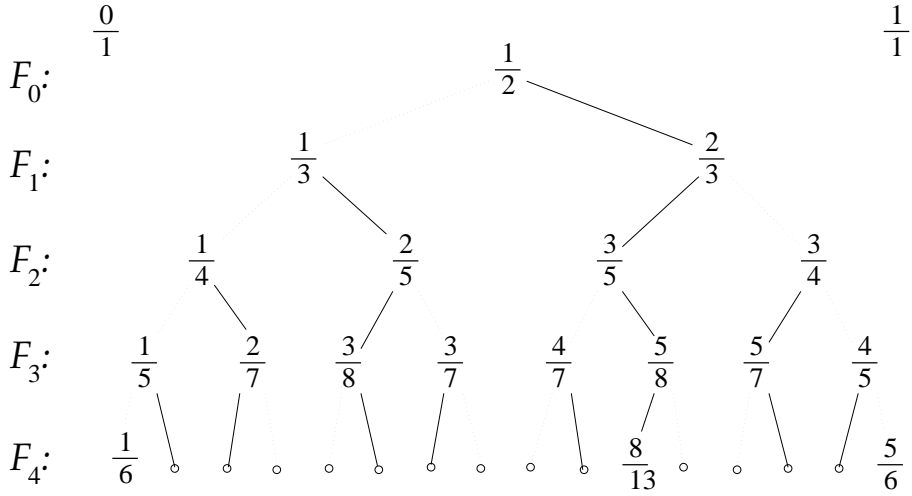,width=12cm}}
\caption[]{
The levels ${\cal F}_n$ of the Farey tree are generated by 
the Farey mediant addition rule $P/Q$, $P'/Q' \to (P+P')/(Q+Q')$.
}
\label{f_farey_sum}
\end{figure}

\begin{figure}
\centerline{\epsfig{file=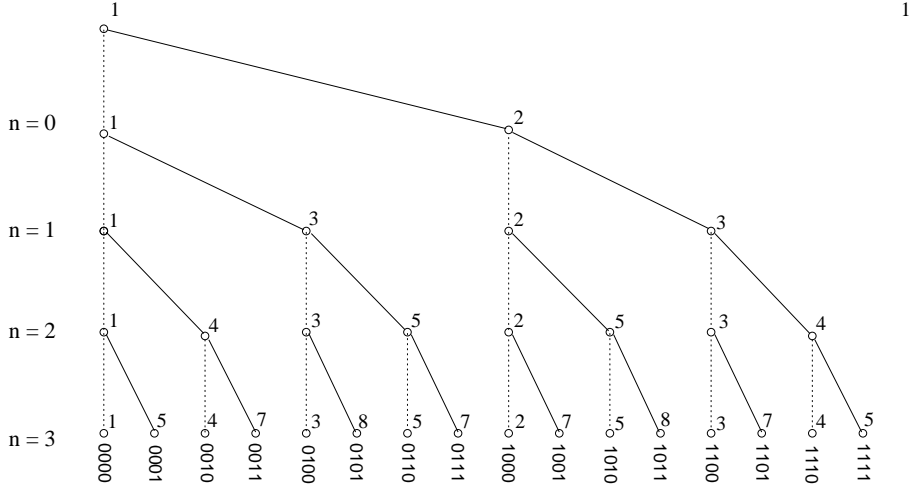,width=12cm}}
\caption[]{
Lexically ordered binary labelling of all Farey
denominators up to and including the $n$th Farey tree level,
following Knauf\rf{Knauf_10}. The binary label of the denominator
$Q_\sigma$ is read off starting with the root of the tree, with
$s_i=1$ for a full line, and $s_i=0$ for a dotted line. For
example $Q_{0011}=7$, $Q_{1000}=2$, etc.. The $2^n$ denominators
$Q_{\cdots 1}$ preceded by a full line constitute the $n$th Farey level 
${\cal F}_n$.
}
\label{f-irrat-Knauf}
\end{figure}

Apply the ``natural measure'' modification of \refeq{weight_L}
\beq
\Lop_{(\tau)}(y,x) =   \left|f'(x)\right|^{\tau} \delta ( x - f^{-1}(y))
\ee{weight_L_Farey}
to the Farey map (we have redefined the exponent $(k+1) \to \tau$
to conform with the notation conventions of \refrefs{chicago5,ACK}), and consider
the sum
\beq
Z_n(\tau) := \int dx \left|{f^n}'(x)\right|^{\tau}  
          \delta(x - f^{-(n+1)}(1)) 
\> .
\ee{averL0x}
The leading $\Lop$ eigenvalue $2^{q(\tau)}$ 
(in the notation of \refref{chicago5}) dominates this sum in the $n\to\infty$ limit
and defines a ``thermodynamic'' function $q=q(\tau)$.
For the Farey map (see \reffig{f_farey_inv}) the pre-images of 1 are
\bea
f({1 / 2}) = 1, 
         \qquad
f^2({1 / 3}) = f^2({2 / 3}) =1
        \continue
f^3({1 / 4}) = f^3({2 / 5}) = \cdots =1
        \continue
f^n({P_i / Q_i}) =  \cdots =1
\,.
\nnu
\eea
These fractions form the levels of 
the {\em Farey tree}\rf{Fareytree,mackay,myrh,CSS,ACK,pres},
a number theoretical construction 
based on the observation that somewhere midway between two
small denominator fractions (such as $1/2$ and $1/3$)
there is
the next smallest denominator fraction (such as $2/5$),
given by the ``Farey mediant'' 
$(P+P')/(Q+Q')$  of the parent mode-lockings
$P/Q$ and $P'/Q'$.
The Farey tree is obtained by starting with the ends of the
unit interval written as 0/1 and 1/1, and then recursively bisecting
intervals by means of Farey mediants. This generates Farey level
sets ${\cal F}_n$ with $2^n$ mode-locking widths on each level, 
\reffig{f_farey_sum}.
%
%
In the present context, the $n$th level of the Farey tree is the set of the $2^n$
distinct backward iterates  $f^{-n}(x_i)=1$ of the Farey map.
Furthermore, noting that $f'_o(x)=1/(1-x)^2$, $f'_1(x)=-1/x^2$, 
it is easily checked that  if $P_i/Q_i \in {\cal F}_n$, then 
\beq
\left|{f^{n}}'(P_i/Q_i)\right| = Q^2_i 
\,.
\ee{(49a)}
For example
\[
{f^{2}}'(1/3) = f'(1/3) f'(1/2) = (1-1/3)^{-2} (1-1/2)^{-2} = 3^2
\]
Hence the sum \refeq{averL0x} is the sum over denominators of the
Farey rationals of the $(n$-1)th Farey level:
\beq  Z_n(\tau)\,=\,\sum_{i \in {\cal F}_n}\,Q_i^{2 \tau}
\,, 
\ee{5.4}
where $Q_i$ is the denominator of the $i$th Farey rational $P_i/Q_i$. For 
example 
\[
  Z_2(1/2)\,=\,4\,+\,5\,+\,5\,+\,4.
\]
As we shall now show, $Z_n(\tau)$ satisfies an exact
sum rule for every non-negative integer $2\tau$.
First one observes that $Z_n(0)=2^n$. It is also easy to check 
that\rf{Fareytree} $Z_n(1/2)=\sum_i Q_i=2\cdot3^n$. More surprisingly, 
$Z_n(3/2)=\sum_i Q^3=54\cdot 7^{n-1}$. Such sum rules are 
consequence of the fact that the denominators on a given level are Farey 
sums of the denominators on preceding levels. In order to exploit
this, the following labelling of the Farey denominators
introduced by Knauf\rf{Knauf_10} is convenient:

The Farey denominators $Q_\sigma$, 
$\sigma \in \{s_1 s_2 \cdots s_n\}$,
$ s_i \in \{0,1\}$, up to the $n$th level of the Farey tree of \reffig{f_farey_sum}
can be labelled in lexical binary order as illustrated in 
\reffig{f-irrat-Knauf}.
%
%
Farey denominators on the $(n+1)$th Farey tree level 
are given by\rf{Knauf_10}
\bea
Q_{\sigma 0} &=& Q_\sigma
        \quad\quad\quad \mbox{~(carry over all previous Farey denominators)}
        \continue
Q_{\sigma 1} &=&        Q_\sigma + Q_{\overline{\sigma}}
        \quad \mbox{(compute the $(n+1)$th Farey tree level)}
\,,
\label{Knauf_comp}
\eea
where $\overline{\sigma}$ is obtained by exchanging 1's and 0's in the string $\sigma$,
$s_i \to 1-s_i$, and the recursion is initiated with the
empty string values $Q_{.}=Q_{\overline{.}}=1$ .
The second rule follows from labelling of nearest
neighbours on a given level, see \reffig{f-irrat-Knauf}.
By construction, $Q_{\sigma 1}$ is invariant under the $s_i \to 1-s_i$
interchange
\beq
Q_{\sigma 1} = Q_{\overline{\sigma}1}
\ee{Knauf_symm}
Define a $[2\tau+1 ]$-dimensional vector 
$\phi = [\phi_0, \phi_1, \cdots, \phi_{2\tau}]$,
$2\tau$ a positive integer, by
\[
\phi^{(k)}_m := \sum_{\sigma \in \{s_1 s_2 \cdots s_k \} }
        Q^{2\tau-m}_{\sigma 1} Q^{m}_{\sigma 0}
        \,=\,\sum_{i \in {\cal F}_{k+1}}\,\left({P_i \over Q_i}\right)^mQ_i^{2 \tau}
\,.
\] 
The zeroth component of $\phi^{(k)}$ is the $k$th level Farey sum
\refeq{5.4}
\beq
Z_k(\tau)\,=\, \phi^{(k)}_0 = \sum_{\sigma \in \{s_1 s_2 \cdots s_k \} }
        Q^{2\tau}_{\sigma 1} 
\,.
\ee{CK_zero}
Motivation for constructing the vector $\phi^{(k)}$ is the observation
that its remaining entries exhaust
all combinations that arise in exponentiating 
$Q_{\sigma 1} = Q_\sigma + Q_{\overline{\sigma}}$ to the integer
powers $s\leq 2 \tau$
\bea
\phi^{(k+1)}_m &=& \sum_{\sigma \in \{s_1 s_2 \cdots s_{k+1} \} }
        \left(Q_{\sigma } + Q_{\overline{\sigma}} \right)^{2\tau-m}
                Q^{m}_{\sigma}
                \continue
     &=& \sum_{\sigma \in \{s_1 s_2 \cdots s_{k} \} }
        \left(Q_{\sigma 1} + Q_{\overline{\sigma}0} \right)^{2\tau-m}
                Q^{m}_{\sigma 1}
        +\left(Q_{\sigma 0} + Q_{\overline{\sigma}1} \right)^{2\tau-m}
                Q^{m}_{\sigma 0}
                \continue
     &=& \sum_{\sigma \in \{s_1 s_2 \cdots s_{k} \} }
        \left(Q_{\sigma 1} + Q_{\sigma 0} \right)^{2\tau-m}
        \left(Q^{m}_{\sigma 1} + Q^{m}_{\overline{\sigma} 1}\right)
\,,
\nnu 
\eea
(we have used $\sum_{\sigma} = \sum_{\overline{\sigma}}$ and the
$Q_{\sigma 1}=Q_{\overline{\sigma}1}$ symmetry), 
and that they form a linear basis for $\phi^{(k+1)}$:
\bea
\phi^{(k+1)}_m
     &=& \sum^{2\tau-m}_{r=0} \combinatorial{2\tau-m}{r}
        \sum_{\sigma \in \{s_1 s_2 \cdots s_{k} \} }
        \left(Q_{\sigma 1}^{2\tau-r} Q_{\sigma 0}^r
              + Q_{\sigma 1}^{r} Q_{\sigma 0}^{2\tau-r} \right)
                \continue
     &=& \sum^{2\tau-m}_{r=0} 
        \left\{ 
          \combinatorial{2\tau-m}{r}+\combinatorial{2\tau-m}{r-m}
               \right\}
                \phi^{(k)}_r
\,.
\nnu 
\eea
Hence the family of vectors $\phi^{(1)}$, $\phi^{(2)}$, $\cdots$,
is generated by multiplication
$\phi^{(k+1)} = {\bf L}_{(\tau)} \phi^{(k)}$ by the 
$[(2\tau+ 1) \times (2\tau+1)]$ dimensional transfer matrix
\beq
{\bf L}_{(2\tau)\,mr} = \combinatorial{2\tau-m}{r}+\combinatorial{2\tau-m}{r-m}
\ee{CK_transfer}
consisting of two Pascal triangles. For example
\[
{\bf L}_{(2)} \,=\, \MatrixIII{2}{4}{2}{1}{2}{1}{1}{0}{1}
\,.
\]
The growth of $\phi^{(k)}$ in the $k\to\infty$ limit (and in
particular, its zeroth component \refeq{CK_zero}, the thermodynamic sum
$Z_k(\tau)$) is given by $1/z=2^{q(\tau)}$,
the leading zero of the ${\bf L}_{(2\tau)}$ characteristic
polynomial
\beq
F(z,2\tau) := \det\left(1- z {\bf L}_{(2\tau)} \right)
\,=\, 1- ( 2\tau+1+F_{2\tau+1}) z - \dots
\,,
\ee{Farey62a}
see \reftab{t_irrat_CIRC}.
Here $F_n=F_{n-1}+F_{n-2}$, $F_0=0$, 
$F_1=1$, are the Fibonacci numbers. The largest $Q_i$ in the $n$th
level sum \refeq{5.4} is $F_{n+2}$, 
so for the large $\tau$ the leading eigenvalue tends to $\rho^{2\tau}$.
The polynomials of \reftab{t_irrat_CIRC}
were first computed by Cvitanovi\'c and Kennedy,
Appendix~B of \refref{ACK}; here we have followed the
more elegant derivation due to 
Contucci and Knauf\rf{CK95}.
%
\begin{table}
\caption[]{
Spectral determinants for the Farey model transfer operator 
\refeq{CK_transfer},
$2\tau=0,\,1,\,2,\dots,11$, together with the leading 
eigenvalue $2^{q(\tau)}$.
        }
\vskip 5pt
\begin{center}
{\small
\begin{indented}\item[]
\begin{tabular}{@{}rcr}
\br
$ 2\tau$ & $2^{q(\tau)}$  & $F(z,\tau)$~~~~~~~~~~~~~~~~ \\
\hline
0 & 2 & $1 - 2\,z $ \\
1 & 3 & $1 - 3\,z $ \\
2 &$ (2.13578\cdots)^{2~}$  & $  1 - 5\,z + 2\,{z^2} $ \\
3 & 7 & $  1 - 7\,z $ \\
4 & $ (1.81346\cdots)^{4~}$ 
      & $  \left( 1 + z \right) \,\left( 1 - 11\,z + 2\,{z^2} \right) $ \\
5 & $(1.75813\cdots)^{5~}$ & $  1 - 14\,z - 47\,{z^2} $ \\
6 & $(1.72342\cdots)^{6~}$ 
            & $  1 - 20\,z - 161\,{z^2} - 40\,{z^3} + 4\,{z^4} $ \\
7 & $(1.69991\cdots)^{7~}$ 
            & $  1 - 29\,z - 485\,{z^2} - 327\,{z^3} $ \\
8 & $(1.68313\cdots)^{8~}$
      & $  \left( 1 + z \right) \,\left( 1 - 44\,z - 1313\,{z^2} - 88\,{z^3} +
       4\,{z^4} \right) $ \\
9 & $(1.67068\cdots)^{9~}$
      & $  1 - 65\,z - 3653\,{z^2} - 3843\,{z^3} $ \\
10 & $(1.66117\cdots)^{10}$
      & $  \left( 1 + z \right) \,\left( 1 - z \right) \,
     \left( 1 - 100\,z - 9601\,{z^2} - 200\,{z^3} + 4\,{z^4} \right) $ \\
11 & $(1.65375\cdots)^{11}$
      & $  1 - 156\,z - 24882\,{z^2} + 83828\,{z^3} + 107529\,{z^4} $ \\
12 & $(1.64784\cdots)^{12}$
      & $ \left( 1 + z \right) \,\left( 1 - 247\,z + 63659\,{z^2} 
          + 797003\,{z^3} - 127318\,{z^4} - 988\,{z^5} + 8\,{z^6} \right) $ \\
$\cdots$ & $\cdots$ & $\cdots$\\
22 & $(1.62413\cdots)^{22}$ & $\cdots$ \\
$\cdots$ & $\cdots$ & $\cdots$\\
$n$ & $\rho^n$ & in the $\to\infty$ limit, 
          $ \quad \rho\,=\,{(1+\sqrt{5}) /{2}} \,=\,$1.61803\dots \\
\br
\end{tabular}
\end{indented}
}  
\end{center}
\label{t_irrat_CIRC}
\end{table}

The Farey model demonstrates that not only are the methods
of this paper competitive with the periodic orbits methods,
but they are sometimes superior to them. 
For the  Farey model the periodic orbit expansions are studied in
detail in \refrefs{may90,AACII}, and depend on 
quadratic irrationals rather than the Farey denominators.
While such periodic orbit expansions are analytically
intractable, this variant of the finite level sums
yields $q(\tau)$ {\em exactly} for all $2\tau$
a non-negative integer. Not only that, but
Contucci and Knauf\rf{CK95} have in this case been
also able to analytically continue the ${\bf L}_{(\tau)}$ 
matrices to arbitrary real positive $\tau $.

\section{Some speculations}

So far we have derived a series of exact sum rules for very specific
simple dynamical systems. In this section we speculate 
about possible, but
at present still largely unexplored 
applications of the above methods in more general settings.

\subsection{Normal form approximations to flows}
\label{s_Norm_forms}

The Poincar\'e maps of flows of physical interest are in
general not polynomials, but smooth analytic functions.
However, in practical applications (such as the long
term integrations in celestial mechanics)  
it is often advantageous to replace a differential equation by
a normal form that approximates the return map for a
Poincar\'e section of the flow, and replace the numerically
demanding integration of the flow by a map iteration.
Such applications require a test of the accuracy of
the normal form approximation; the above sum rules offer an estimate of
the quality of approximation, which, as it depends only on 
cycle eigenvalues, is coordinatization independent. The idea
is simple: were the polynomial representation of the flow
exact, its cycles would satisfy the above sum
rules exactly. By comparing the sums of cycles
of the smooth flow to what would be expected were the 
flow generated by a polynomial we learn how good a polynomial
approximation of given order would be, prior to any 
actual fitting of the flow by an approximate mapping. 

%
As an example, consider
a 3-dimensional flow
$\dot{\bf x}={\bf F}({\bf x}),\;\; {\bf x}=(x_1,x_2,x_3)$.
We assume that the flow of interest is recurrent,
and that given a convenient Poincar\'e section coordinatized by
coordinate pair $(x,y)$, the flow can be described
by a 2-dimensional Poincar\'e map
\beq
\left. {
         x' = f(x,y)
         \atop
         y' = g(x,y)
         } \right.
\,, 
\ee{Section}
together with the ``ceiling" function $T(x,y)$ which gives the 
time of flight to the next section for a trajectory starting at $(x,y)$.
In general $f$ and $g$ can be complicated functions, but the 
essential properties of a continuous flow can be 
modelled by the H\'enon map \refeq{henon_m}, which we take as a local 
normal form (up to quadratic terms). 
In the Hamiltonian case, the normal form is
\beq
x_{k+1} = 1-ax^2_k-x_{k-1}
\,.
\label{time_spread}
\eeq

If this
were an exact representation of the flow, it would 
satisfy exactly the sum rules 
\refeq{1st-2dHam} and \refeq{2nd-2dHam},
hence the amount by which they are violated is an indication of how well
the flow is approximated by a quadratic normal form. 
As we are summing signed rather than absolute values,
this is an underestimate of the actual error, but possibly of
interest even so.
While numerical tests of the rule on a 3-disk billiard cycles data
of \refref{CRR93} do suggest that for sufficiently large
disk-disk separation \refeq{time_spread} is not a bad approximation,
we have not yet checked
whether the best value of the parameters
in the approximation can be determined by minimizing the deviation
from the exact sum rules.

\subsection{Spectrum of the period-doubling operator}
\label{s_period-doub}

The generalized \evOper s \refeq{weight_L} have an 
important application\rf{AACI,Jiang,CCR,Pollicott91}
in the period-doubling renormalization theory:
if $f$ is given by the period-doubling presentation function\rf{pres}
\[
\begin{array}{lccl}
f_0 (x) &=&
\alpha g (x), \quad & g \left(\alpha^{-1}\right) \leq x \leq 1 \\
f_1 (x) &=&
\alpha x, \quad & \alpha^{-1} \leq x \leq \alpha^{-2} \,\, ,
\end{array}
\]
where
\[
g(x)=\alpha g \circ g (x/\alpha ) \,\, .
\]
is the universal period--doubling renormalization fixed point
function, the leading eigenvalue of 
$ \Lop_{(2)} (y,x) =  \delta ( y - f^{-1}(x))f'(y) $ is
the Feigenbaum $\delta$, and the spectrum of $ \Lop_{(2)}$ is given
by the trace formula\rf{AACI,CCR}
\beq
\tr \Lop_{(2)} = 
 \sum_{i\inFix{}} \frac{\Lambda_i^2}{\Lambda_i-1} 
\,,
\label{p.d_sr}
\eeq
or the associated \Fd.
Here the summation is restricted to the real fixed points in the
unit interval; application of our sum rule \refeq{2nd_sr}
would require inclusion of the complex periodic points as well.
The crudest approximation to $g(x)$ is a quadratic 
polynomial. This already yields a reasonable estimate without any
actual polynomial fitting; 
according to \refeq{2nd_sr}, for a
quadratic polynomial the leading eigenvalue \refeq{2nd_sr} is 4, 
while the period--doubling
leading eigenvalue is the Feigenbaum $\delta = 4.66\dots$ . If
\refeq{2nd_sr} is really applicable in this context, the sum rule would 
require inclusion of contributions of
complex periodic points of the presentation function -
how well that converges has not yet been checked.

\section{A critical summary}

The conventional periodic orbit theory suffers from one serious
limitation; the number of spectral eigenvalues computable is in
practice limited by the exponential growth  in the number of
periodic orbits required. This arises because the periodic orbit
theory in its trace formula formulation is essentially a tessellation
of the phase space by linearized neighbourhoods of periodic points which
ignores the fact that these neighbourhoods are interrelated by the
analyticity of the flow. This suggests that the local periodic orbit
information should be supplemented by global analyticity constraints.
Previous to the work presented here, analyticity has been exploited
to improve the convergence of cycle 
expansions by incorporating the shadowing of long cycles by
combinations of shorter ones\rf{cycprl,AACI}, and to prove that
the cycle expansions of \Fd s can converge 
faster than exponentially with the 
topological cycle truncation length\rf{grothi,Rugh92}.
While sometimes very useful, these results apply only to 
nice hyperbolic flows with finite symbolic dynamics.
Most realistic physical flows are not uniformly hyperbolic, and
do not have finite Markov partitions. 
For this reason general analyticity constraints applicable to
any flow would be of great interest.  Here another observation
is very suggestive; analytic continuation of pairs
of orbits through inverse bifurcations\rf{KHD}
improves the semi-classical quantization in a regime where the dynamics
is non-hyperbolic, and utility of periodic orbits {\em a priori}
in doubt. 
Similarly, leading 
diffraction effects of wave mechanics can be accounted for
by inclusion of complex ``creeping'' orbits\rf{VWR} into periodic
orbits sums.
These examples suggest that for generic flows 
perhaps both the non-hyperbolicity and 
lack of finite Markov partitions may be tamed by
taking into account the totality of periodic orbits, real and
complex.

With such considerations in mind we have here compared two strategies
to compute the spectrum of a given \evOper :
\begin{itemize}
\item[(i)]
the conventional periodic orbit theory approach: $\tr \Lop =$ sum over cycles
\item[(ii)]
the new recursive approach: relate
$Z_n = \int dx\, \Lop^n(0,x) = $ sum over pre-images to previous levels sums.
\end{itemize}
We have established that sometimes the iterative nature of the dynamics
enables us to obtain formulas for \evOper\ eigenvalues without
computing {\em any} periodic orbits.

The sum rules so obtained come as something of a pleasant surprise,
especially in the multi-dimensional cases, as cycle expansions
typically require evaluation of exponentially many cycles up to
a given cycle length, and it is not at all obvious that
the cycle weights should be related in any simple way. 
Prior to these results, 
other than the purely topological Lefschetz fixed point formulas
there existed only one exact general result
for periodic orbit sums, the flow density conservation\rf{HOdA84,AACI},
and several sum-rules specific to billiards. These are explored
in a forthcoming paper by S.F.Nielsen et al.\rf{sune}.
From the vantage point of this work, the flow density conservation rule 
is of limited interest, as it has no dynamical content.
Even though our sum rules
require determining not only real cycles, but also the
complex ones ({\em ie.} also those forbidden by the symbolic dynamics),
they might be useful in practice. There are
several immediate uses for such rules; one is to 
check for the correctness of the numerically calculated
cycle weights up to a given cycle length, and another, speculated on
in \refsect{s_Norm_forms}, is to estimate the quality of
polynomial approximations to smooth flows.

Applied to higher dimensional dynamical systems 
the new sum rules might shed light on the role of complex
periodic orbits in phenomena such as intermittency and
pruning, reduce the number of periodic orbits required for
accurate eigenspectra computations, provide useful 
cross checks of the accuracy of  trace formulas and cycle expansions, 
might lead to replacement of periodic orbit sums by contour integrals,
and provide  error bounds on approximations of smooth flows by 
normal form mappings.
Still, the explicit sum rules for simple models presented above
are only a modest step in direction of
moving ``beyond periodic orbit theory'':
\begin{itemize}
\item[(a)]
        The results presented
        depend on either the finiteness of polynomials
        (for the contour integral method) or properties of integers (for the
        Farey map case).
        One needs to develop infinite dimensional versions of the 
        above finite rank operators
        ${\bf L}$ (\refref{CK95} is a step in this direction).
\item[(b)]
        One needs to reinstate positive measures. While in quantum mechanical
        applications the signed measures might be just the right Maslov phases,
        in the classical context the above sum rules are essentially index
        theorems and play a role analogous to what Lefschetz zeta function
        plays vis-a-vis the Artin-Mazur counting zeta function.
        (Appendix of \refref{Levin93} is a step in this direction).
\item[(c)]
        The \evOper\ weights studied here were restricted to
        powers of the cycle stability; one needs to 
        deal with more general classes of weighted operators.
\end{itemize}

\ack
Authors are grateful to HH~Rugh
for bringing \refref{Levin91} to their attention, 
G Anderson\rf{Ander_95} for kindly sending us his unpublished notes with
a different proof of the rationality of the Hatjispiros-Vivaldi zeta functions,
and to G~Tanner for a critical reading of an early version of the manuscript. 
The transfer matrices \refeq{B_lige} for the quadratic
case were derived by KH as a part of the Niels Bohr Institute 3rd year 
undergraduate physics project. 
PC is grateful to the Isaac Newton Institute, Cambridge, and Dj.~Cvitanovi\'c,
Kostrena, where significant
 parts of this work was done,
for kind hospitality, and the Carlsberg Fundation for many years of unflagging
support. JR gratefully acknowledges financial support by the Studienstiftung
des deutschen Volkes.
GV thanks the support of the Hungarian Science Foundation
OTKA (F019266/F17166/T17493) and the Hungarian 
Ministry of Culture and Education FKFP 0159/1997.

\appendix
\section{Examples of sum rules}

To give the reader a feeling for what the above sum rules mean in practice,
we shall work out explicitly the few simplest examples.
The quadratic polynomial fixed points and their stabilities are given by 
\bea
x_0 &=& (1 + \sqrt{1+4c})/2\,, \quad
x_1 = (1 - \sqrt{1+4c})/2\,, \quad
                \continue
\Lambda_0 &=& 2 x_0 \,, \quad \Lambda_1 = 2 x_1
\,,
\eea
and the 2-cycle stability by
\[
\Lambda_{01} = 4(1-c)
\,.
\]
\subsection{Trace formula for $\Lop_{(0)}$}

The sum rule \refeq{tr_0} states that for 
the quadratic polynomial \refeq{fateau}
\[
0= {1 \over \Lambda_0-1} + {1 \over \Lambda_1-1}
\,, \quad
0= {1 \over \Lambda_0^2-1} + {1 \over \Lambda_1^2-1}
+ {2 \over \Lambda_{01}-1}
\,,
\]
and so on.
This sum rule is not new; it 
goes back to Fatou\rf{fatou} and Julia\rf{julia} and
can be found, for example, in sect.~3.5 of
\refref{deva87}, proven by manipulating roots of polynomials
rather than by the contour integration method.
The \Fd\ \refeq{Fd:0:1:2} is in this case trivial,
$
F{(z,0)} =1
$.

\subsection{Trace formula for $\Lop_{(1)}$}

The sum rule \refeq{1st_sr}  for the quadratic polynomial 
fixed points and 2-cycles can be checked by hand:
\[
2= {\Lambda_0 \over \Lambda_0-1} + {\Lambda_1 \over \Lambda_1-1}
\,, \quad\quad
4= {\Lambda_0^2 \over \Lambda_0^2-1} + {\Lambda_1^2 \over \Lambda_1^2-1}
+ 2 { \Lambda_{01} \over \Lambda_{01}-1}
\,.
\]
For quadratic polynomials $N = 2^n$, so both
the \Fd\ \refeq{Fd:0:1:2} and the \dzeta\ \refeq{zeta_k} are given by
$
F{(z,1)} =1-2z
$,
$
1/\zeta_{(1)} = { F{(z,1)} / F{(z,0)}} = 1-2z
 \,\, .
$ 
For the case of all periodic points real, the traces $T_n(1)$
count the numbers of periodic points of period $n$, and
$1/\zeta_{(1)}$ is the Artin-Mazur
topological zeta function\rf{AM}.

\subsection{Trace formula for $\Lop_{(2)}$}

The sum rule \refeq{2nd_sr}  for the quadratic polynomial
is given by
\[
 \sum_{i\inFix{n}} \frac{\Lambda_i^2}{\Lambda_i-1} = 4^n
\,,
\]
the \Fd\ by 
$
F{(z,2)}(z) =1-4z
$, 
and the \dzeta\ by      
\[
1/\zeta_{(2)} = { F{(z,2)} \over F{(z,1)}} = {1-4z \over 1-2z}
 \,\, .
\] 
Furthermore, as for every $k>0$
the term \refeq{contour5} contributes a factor $(1-2^{k}z)$
to $ F{(z,k)}$, $1/\zeta_{(k)}$ always contains a factor
$(1-2^{k}z)/(1-2^{k-1}z)$, in agreement with \refref{HV93}.

\subsection{Trace formula for $\Lop_{(3)}$}

According to \refeq{12a}, we only need to
to evaluate $C{(3)}$ for the quadratic polynomial 
\[ 
C{(3)} =  \Cint{x} {8x^3 \over x^2 -c} =
   8 \Cint{x} x (1 + {c \over x^2} + \dots) = 8c
\,.
\]
Together with \refeq{3rd_sr}
this yields the $T_n(3)$ sum rule for the quadratic polynomial
\beq
 \sum_{i\inFix{n}} \frac{\Lambda_i^3}{\Lambda_i-1} = 8^n + (8c)^n
\,.
\ee{3rd_sr_quad}
For example, it is easily verified that
\bea
8(1+c)&=& {\Lambda_0^3 \over \Lambda_0-1} + {\Lambda_1^3 \over \Lambda_1-1}
\continue
64(1+c^2)&=& {\Lambda_0^6 \over \Lambda_0^2-1} + {\Lambda_0^6 \over \Lambda_1^2-1}
+ 2 {\Lambda_{01}^3 \over \Lambda_{01}-1}
\,.
\nnu
\eea

The \Fd\ is given by
\[
F{(z,3)}(z) =(1-8z)(1-8cz)
\,\,,
\] 
in agreement with 
the $k=3$ case of our general formula \refeq{Fred_B}. The \dzeta\ is
\beq
1/\zeta_{(3)} = { F{(z,3)} \over F{(z,2)}} = {(1-8z)(1-8cz) \over 1-4z}
 \,\, .
\ee{zeta_3}
Thus the first ``nontrivial'' zeta function is indeed of the
form conjectured in \refref{HV93}.

\subsection{Trace formula for $\Lop_{(7)}$}
\label{s_Tr_L_7}

We finish this appendix by writing out explicitely 
a typical nontrivial example: $k = 7 = 2 \times 3 + 1$.
According to \refeq{B_lige}
\[
{\bf L}_{(7)} = 
\left[ 
\matrix{ {c^3} & 3\,c & 0 \cr {c^4} & 6\,{c^2} & 1 \cr {c^5} & 10\,{c^3} & 
5\,c \cr  }
\right] 
\,.
\]
The corresponding \Fd\  is given in \reftab{t_quadr_map}.
This, as well as any other $F(z,k)$ that we have computed
agrees with the form conjectured in \refref{HV93}.

%
%
%
%

\section*{References} 

\end{document}